\documentclass[slac_one]{revtex4}
\usepackage{graphicx,epsfig}
\usepackage{fancyhdr}
\begin{document}

\title{Electroweak Cross-sections and Widths} 

%

\author{A. Robson (on behalf of the CDF and D0 Collaborations)}
\affiliation{Glasgow University, Glasgow G12 8QQ, UK}

\begin{abstract}
The status of W and Z cross-section and width measurements 
from the CDF and D0 experiments is reviewed.
Recent results that are discussed: the 
cross-section for Z production times 
the branching ratio to tau pairs, the rapidity and transverse 
momentum distributions of Z production in the electron channel, 
and the direct measurements of the W width and the Z invisible 
width; the latter from an analysis of events with large missing 
transverse energy and one or more energetic jets.
\end{abstract}

\maketitle

\thispagestyle{fancy}


\section{W AND Z PHYSICS AT THE TEVATRON} 

Already early in Run 2 of the Tevatron, using less than 
100\,pb$^{-1}$ of data, total inclusive 
W and Z cross-section measurements were 
essentially systematically limited \cite{ref:cdfwzprl}.
From the theoretical side, W and Z cross-sections are 
well-known, fully differentially to NNLO.  
From an experimental point of view, the Tevatron W and Z boson datasets 
provide a pure sample of high-$p_{\rm T}$ electrons and muons.  
Now with more than thirty times the statistics of those early 
measurements, the W and Z datasets provide a powerful tool 
for lepton reconstruction, identification and trigger measurements. 
But at a fundamental level, total inclusive W and Z cross-section 
measurements are benchmarks for all high-$p_{\rm T}$ physics analyses.

A tower of physics measurements is built on the selection 
of W and Z events. W and Z plus photon measurements and heavy 
diboson WW, WZ and ZZ measurements probe triple 
gauge couplings; 
and the precision of top-quark cross-sections continues to improve.
Each of these measurements needs to demonstrate consistent 
determinations of inclusive W and Z cross-sections as a test of 
the implementation of selections, efficiencies, good run lists 
and luminosity computations.
Observations of the rarer processes can be seen 
as stations on the way to a heavy Higgs boson, decaying to a 
pair of W bosons \cite{ref:higgs}.

However dedicated measurements on single W and Z events also continue.
Their focus now is on the more challenging 
tau decay modes, and on using the large datasets to make 
differential cross-section measurements in order to look for 
discrepancies with higher-order calculations and to test the 
edges of phase space, and to make high-precision measurements 
of Standard Model parameters such as masses and widths.

\section{Z $\rightarrow\tau\tau$}
Tau leptons are more complex objects to reconstruct than electrons or muons, 
and measurement techniques continue to improve.  
At D0, $\tau$ identification starts from calorimeter clusters, 
reconstructed in a cone of $\Delta R<0.5$, that have their energy 
concentrated in an inner cone of $\Delta R<0.3$.  The tracks 
within the inner cone are required to be consistent with a 
tau, such that $m_{\rm tracks}<1.8$\,GeV.  
The electromagnetic clusters are considered in terms of sub-clusters, 
reconstructed from a seed in the most finely segmented `shower-maximum' 
layer of the calorimeter.

Tau candidates are assigned to one of three categories, according 
to the number of tracks and subclusters.  The categories correspond 
broadly to one-prong, one-prong plus neutral, and three-prong tau decays.

To improve the selection, neural nets are trained for each category on 
variables that characterise the isolation, shower shape, and correlations 
between the tracks and clusters.  This results in excellent 
background suppression.

Furthermore, recent improvements in the $\tau$ energy corrections 
have significantly reduced uncertainties.

D0 selects Z$\rightarrow\tau_\mu\tau_{\rm h}$ from an inclusive muon trigger.
Both the muon and tau have $p_{\rm T}>15$\,GeV, and in addition there 
is a minimum requirement on the scalar sum-$p_{\rm T}$ of the tau tracks.  
The leptons have opposite charge.

A data-driven estimate of the QCD background, which is mostly ${\rm b\overline{b}}$, 
is obtained from same-charge events, and corrected for the expected rate of 
same-charge events from a QCD-enhanced dataset.  Electroweak backgrounds are 
obtained from simulation, and the W plus jets background is corrected 
for the component already accounted for in the same-charge events.
The background uncertainties are much reduced compared to previous D0 measurements.

The resulting selection is shown in Figure\,\ref{fig:Ztautauplot}.  1511 events 
are observed, of which around 20\% are background.  A cross-section is 
extracted:
$\sigma({\rm p\overline{p}\rightarrow Z})\cdot Br({\rm Z\rightarrow\tau\tau}) 
 = 240 \pm 8({\rm stat}) \pm 12({\rm sys}) \pm 15({\rm lumi}) \,\, {\rm pb}$ 
\cite{ref:d0newZtautau}.

\section{Z RAPIDITY}

In Z boson production, the rapidity $y=\frac{1}{2} \ln \frac{E+p_z}{E-p_z}$ 
is closely related to the momentum fractions $x$ of the interacting partons: 
at leading order the relation is $x_{1,2} = \frac{m}{\sqrt{s}}e^{\pm y}$.  
Measuring the Z boson rapidity is therefore a direct probe of the PDFs of 
the interacting partons.  

Furthermore, at CDF, Z bosons can be reconstructed over the entire 
kinematic range in the electron channel, using the forward calorimeters.  
This gives access to events at very low $x$, where PDFs are relatively 
uncertain. 

Events are selected in three topologies, according to whether the electrons 
are reconstructed in the central or forward parts of the detector.  

Backgrounds are estimated in a data-driven way, by forming signal and 
background templates of the electron isolation distribution, and extrapolating 
the non-isolated tail into the low isolation signal region. 

The acceptance as a function of rapidity is taken from simulation, and 
the rapidity dependences of the electron identification efficiencies are determined. 

\begin{figure}[t]
\includegraphics[width=60mm]{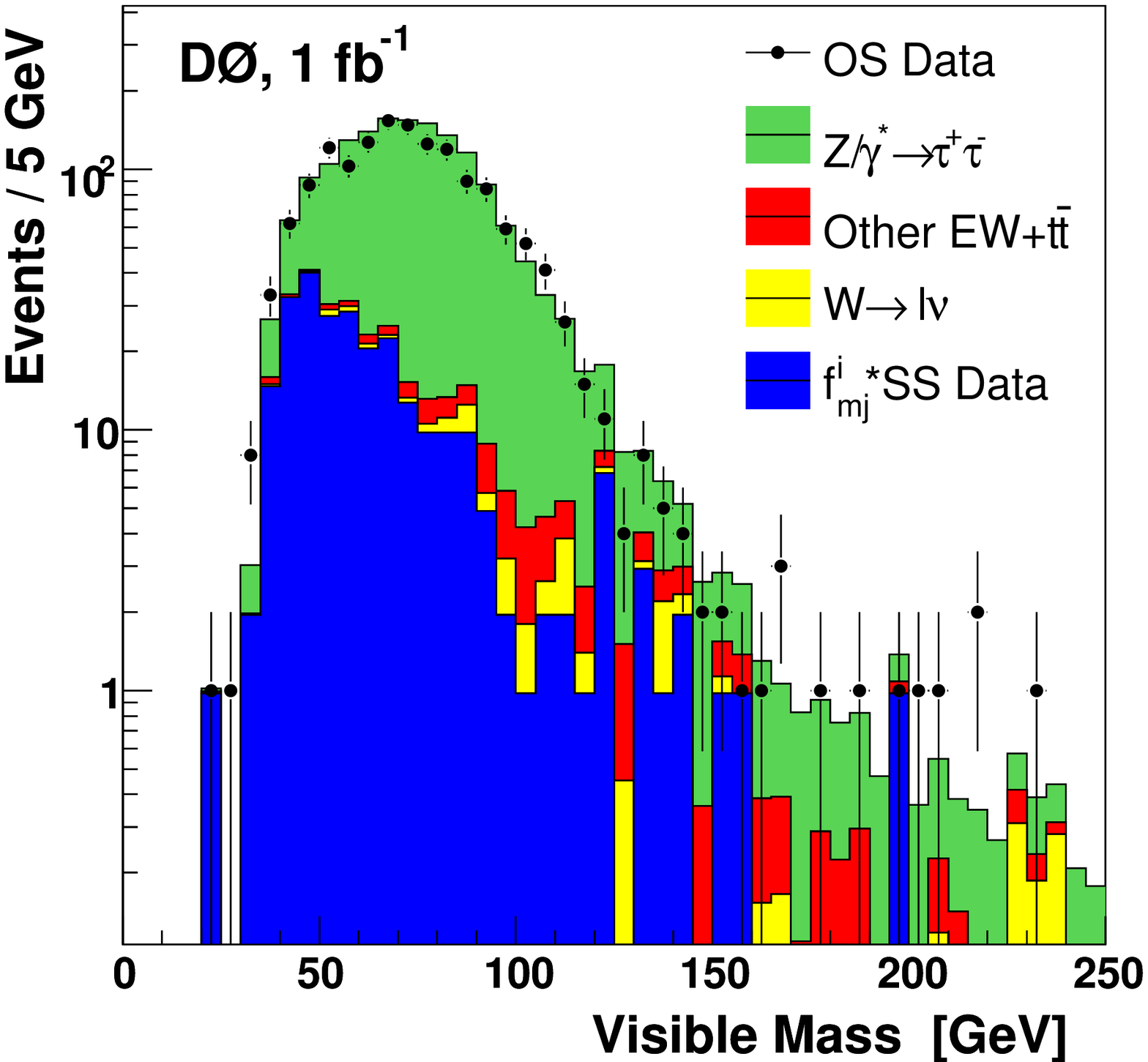}
\includegraphics[width=75mm]{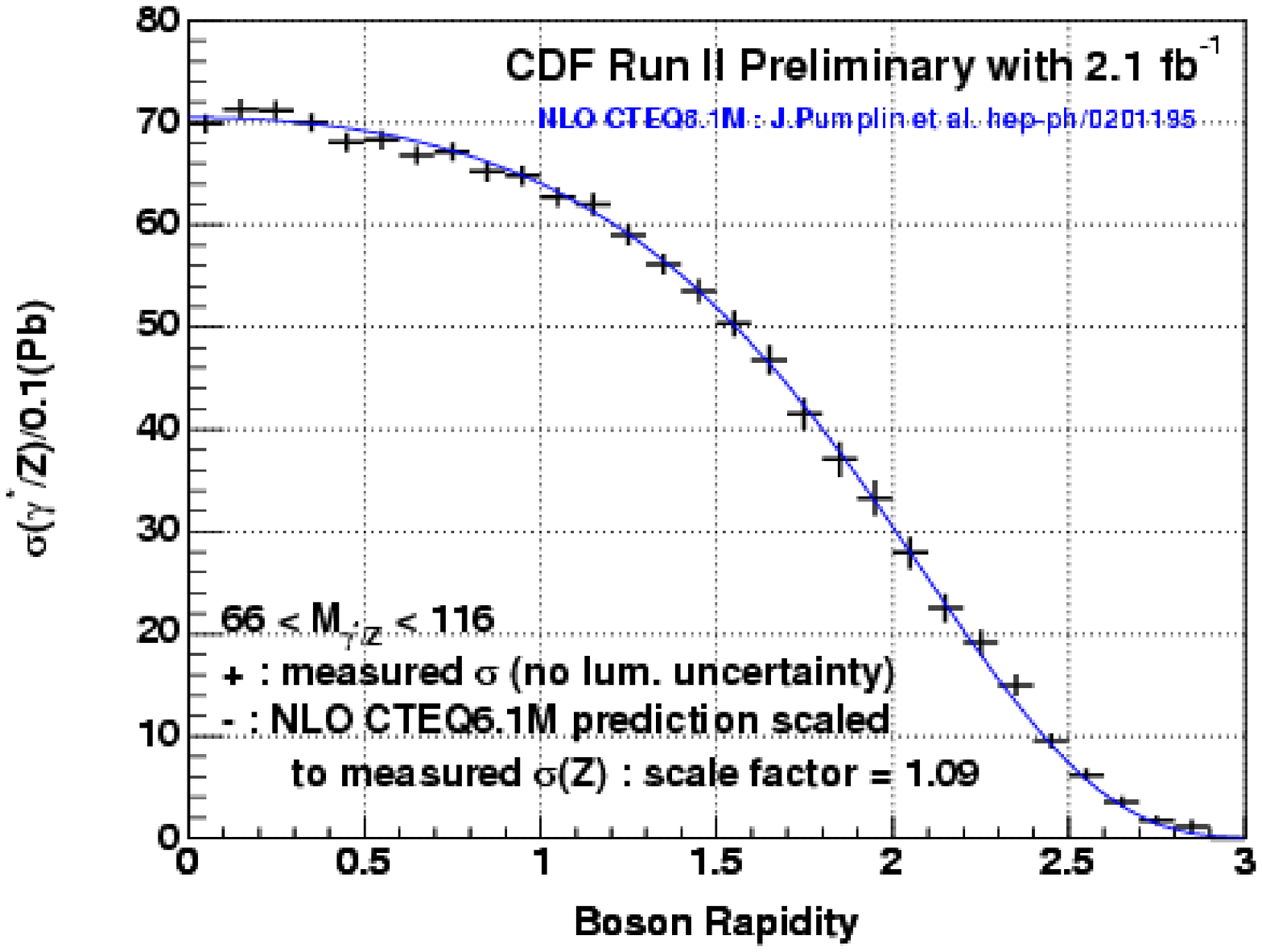}
\caption{(left) The `visible mass', $m_{\rm vis}=\sqrt{(P_{\mu}+P_{\tau}+\not{\!\!\! P_{\rm T}})^2}$, for Z$\rightarrow\tau_\mu\tau_{\rm h}$ candidates.  (right) The Z rapidity distribution.} \label{fig:dsdy}\label{fig:Ztautauplot}
\end{figure}

The measured rapidity distribution is shown in Figure\,\ref{fig:dsdy}.
Whereas previous versions of this measurement have been entirely statistically 
dominated, the current large dataset results in systematic uncertainties 
comparable with the statistical uncertainties in some regions of rapidity. 
Comparison is made with an NNLO calculation and NNLO PDFs, and 
with an NLO calculation and several NLO PDFs. 
Although at present the measurement does not clearly favour one set over another, 
with even more statistics there could be some scope for constraining PDFs.


\section{Z TRANSVERSE MOMENTUM}

Measuring the $p_{\rm T}$ of the Z tests QCD predictions 
for initial state gluon radiation.
Whereas the high end of the $p_{\rm T}$ spectrum (above $\sim$30\,GeV/$c$) 
is dominated by single hard emissions and perturbative QCD is reliable, 
the low end of the spectrum is dominated by multiple soft gluon radiation, 
which must be calculated by resummation or modelled by a non-perturbative 
parton shower monte carlo.  
The {\sc Resbos} event generator 
implements NLO QCD 
and the CSS resummation formalism, using the BNLY form-factor and 
parameters determined by global fits to DIS and fixed-target data.
This is a particularly interesting time to be probing this model, as 
recent global fits have suggested the presence of an extra contribution 
to the form-factor at small $x$.  This would imply a broadening of the 
Z $p_{\rm T}$ at high rapidities ($|y_{\rm Z}|>2$) at the Tevatron, and 
a significant effect on centrally-produced W and Z bosons at the LHC.  
D0 has looked for evidence of this effect \cite{ref:d0pt}.

In 1\,fb$^{-1}$, around 64000 Z events are reconstructed in the electron channel, 
of which around 5000 have $|y_{\rm Z}|>2$.
Backgrounds are estimated from templates fitted to the mass distribution 
$m_{\rm ee}$.  A regularized unfolding technique is used to recontruct 
the underlying $p_{\rm T}$ distribution. 
The low end of the Z $p_{\rm T}$ distribution for all rapidities is compared 
to the {\sc Resbos} prediction, and found to be in good agreement.  
The Z $p_{\rm T}$ for events that have $|y_{\rm Z}|>2$ is 
compared to {\sc Resbos} predictions both with and without the extra 
small-$x$ form-factor, as shown in Figure\,\ref{fig:Zpt}.
Without the small-$x$ form-factor the fit is found to be $\chi^2$/dof=11.1/11, 
whereas with the extra term the fit is $\chi^2$/dof=31.9/11.  
This measurement therefore disfavours the small-$x$ broadening.

It is interesting to note that when the complete Z $p_{\rm T}$ spectrum 
is examined, a NNLO calculation 
reproduces the shape well, but requires rescaling by 25\% to match the normalisation.

\section{W WIDTH}

The width of the W boson is predicted very precisely in the Standard Model, 
and so its accurate measurement is a powerful check of the consistency 
of the Standard Model.  
CDF has measured $\Gamma_{\rm W}$ using 350\,pb$^{-1}$ of data \cite{ref:cdfwwidth}.

Experimentally, transverse quantities are accessible at the Tevatron; for 
example the transverse mass 
$m_{\rm T}=\sqrt{2 p_{\rm T}^{\ell} p_{\rm T}^{\nu} (1-\cos{\phi_{\ell\nu}})}$.
Events can have $m_{\rm T} > m_{\rm W}$ either as a result of the 
intrinsic W width, or as a result of detector 
smearing.  The Gaussian effects of detector smearing are found to 
fall faster than the intrinsic Breit-Wigner lineshape, so the 
procedure for measuring the width is to use the region $m_{\rm T}<90$\,GeV/$c^2$
for normalisation, and to fit templates to the high $m_{\rm T}$ region.

Events to construct the templates are taken from a leading-order monte carlo, 
matched with {\sc Resbos} for QCD initial state radiation and with
a calculation from Berends and Kleiss for QED final state radiation. 
A fast simulation models electron conversion and showering, muon 
energy loss, and includes a parametric model of the energy in the 
detectors not associated with the high-$p_{\rm T}$ electron or 
muon (the `recoil energy'), which originates from QCD, the 
underlying event and bremstrahlung. 
The final uncertainty on $\Gamma_{\rm W}$ from the recoil modeling 
is $\Delta\Gamma=54$\,MeV and $\Delta\Gamma=49$\,MeV in the electron 
and muon channels respectively. 
The modeling of the tracking scale and resolution and the calorimeter 
energy scale and resolution are checked on Z$\rightarrow\mu\mu$ and 
Z$\rightarrow$ee events respectively; and also on the J/$\psi$.  
Uncertainties from the tracking scale and resolution are 
$\Delta\Gamma=17;26$\,MeV (e;$\mu$) and from the calorimeter scale 
and resolution $\Delta\Gamma=21;31$\,MeV (e;$\mu$).
Backgrounds are dominated by QCD multijet events in the electron 
channel and by Z$\rightarrow\mu\mu$ and decays in flight in the 
muon channel ($\Delta\Gamma=32;33$\,MeV (e;$\mu$)).


\begin{figure}[t]
\includegraphics[width=70mm]{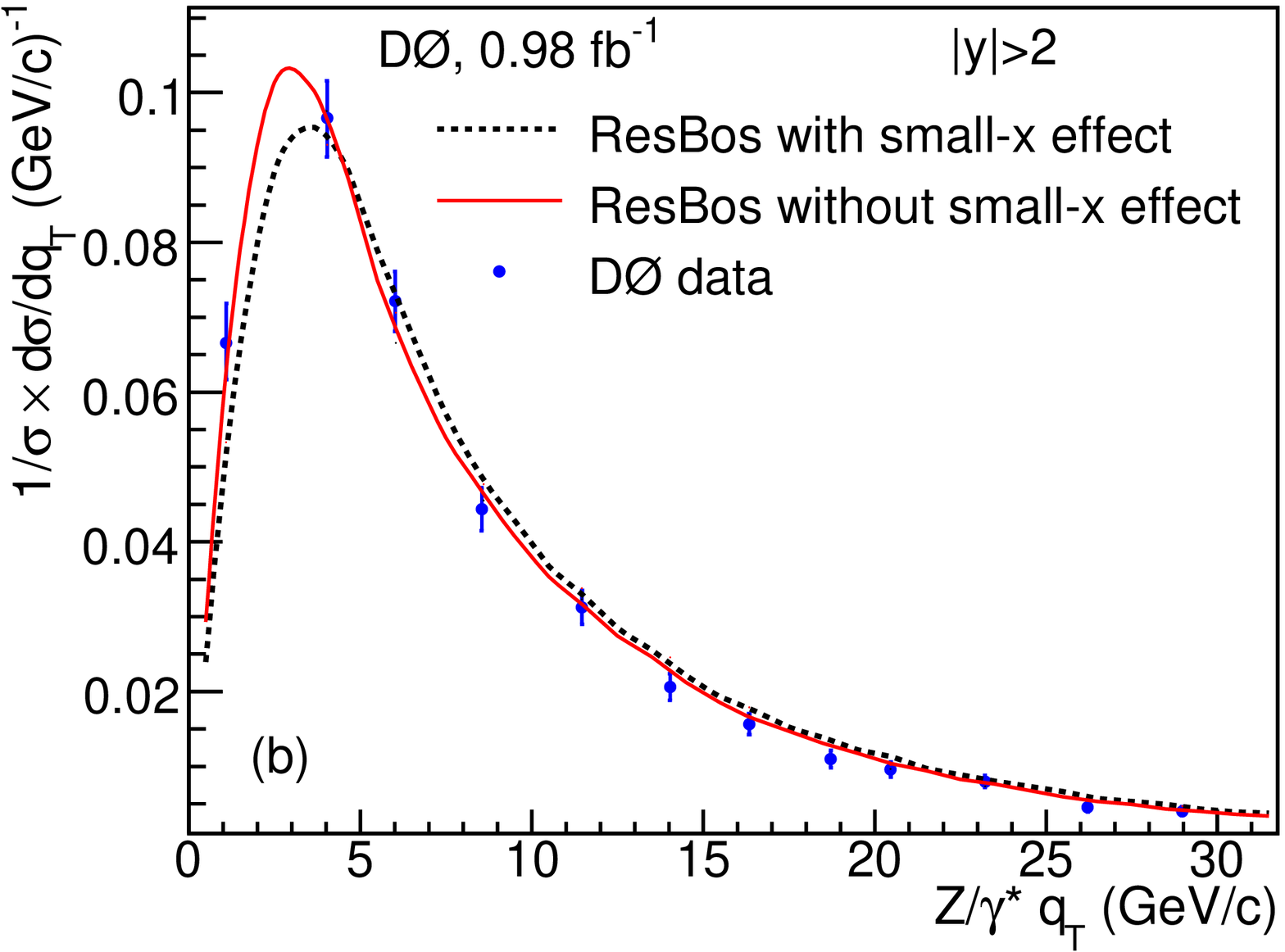}
\includegraphics[width=65mm]{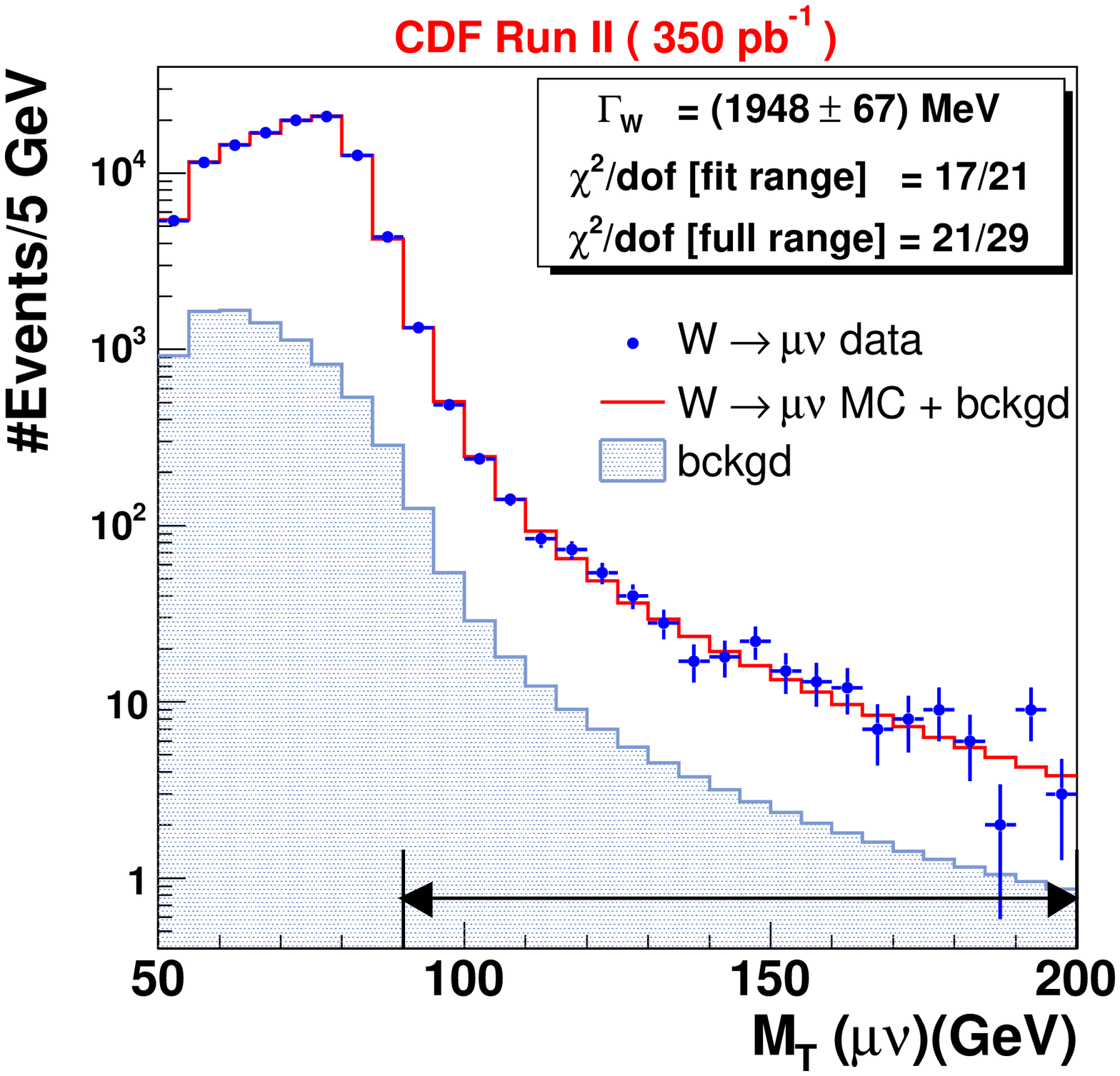}
\caption{(left) The Z $p_{\rm T}$ distribution for high-rapidity events.  (right) The transverse mass fit in the muon channel.} \label{fig:Wwidth}\label{fig:Zpt}
\end{figure}

The fit in the muon channel is shown in Figure\,\ref{fig:Wwidth}.
The fits are excellent and the W width is found to be:
\begin{equation}\label{eq:wwidth}
\Gamma_{\rm W} = 2032 \pm 73 ({\rm stat\,+\,sys})\,\,{\rm MeV} \, ,
\end{equation}
which is the world's most precise single direct measurement.
It is consistent with the Standard Model prediction 
($\Gamma_{\rm W} = 2093 \pm 2$\,MeV) \cite{ref:pete} 
and with CDF's earlier indirect measurement 
($\Gamma_{\rm W}^{\rm indirect} = 2092 \pm 42$\,MeV) \cite{ref:cdfwzprl}.

\section{Z INVISIBLE WIDTH}

The invisible width of the Z is measured very precisely indirectly 
from LEP: $\Gamma_{\rm Z}({\rm invis})=500.8\pm2.6$\,MeV. 
However the LEP direct measurement, using single photon events, 
is much less precise: $\Gamma_{\rm Z}({\rm invis})=503\pm16$\,MeV. 
An analagous measurement from CDF in 1\,fb$^{-1}$ \cite{ref:zinviswidth}, using events with a single 
jet and large missing $E_{\rm T}$, is completely uncorrelated. 

Single-jet events are selected from a missing $E_{\rm T}$ trigger.  
Independently, $\sigma({\rm Z\,+\,1\,jet})\cdot Br({\rm Z\rightarrow\ell\ell})$ 
is measured from high-$p_{\rm T}$ lepton triggers.  
The ratio of invisible and visible widths can be written as the 
ratio of cross-sections:
\begin{equation}\label{eq:Zinviswidth}
\frac{\Gamma_{\rm Z}^{\rm invis}}{\Gamma_{\rm Z}^{\ell\ell}}
= \frac{\sigma({\rm Z\,+\,1\,jet})\cdot Br({\rm Z\rightarrow invis})}{\sigma({\rm Z\,+\,1\,jet})\cdot Br({\rm Z\rightarrow\ell\ell})}
= \frac{N_{\rm obs} - N_{\rm bck}}{\kappa\cdot\sigma({\rm Z\,+\,1\,jet})\cdot Br({\rm Z\rightarrow\ell\ell})\cdot\mathcal{L}} \,\,\, ,
\end{equation}
where $\kappa$ is a correction to take into account the different 
acceptance of the `+1\,jet' selection in Z$\rightarrow\nu\nu$ 
events compared to Z$\rightarrow\ell\ell$ events.

The visible lepton cross-section is measured to be 
$\sigma({\rm Z\,+\,1\,jet})\cdot Br({\rm Z\rightarrow\ell\ell})=0.555\pm0.024$\,pb. 
This leads to an extracted value $\Gamma_{\rm Z}^{\rm inv} = 466 \pm 42\,\,{\rm MeV}$,
where the electroweak backgrounds, QCD backgrounds and visible lepton 
cross-section make approximately equal contributions to the uncertainty.
This can also be interpreted as a measurement of the number of neutrino 
species, $N_{\nu}=2.79\pm0.25$.

With more data this measurement could become competitive with 
those from the LEP experiments.

\section{CONCLUSIONS}

W and Z cross-section measurements underpin the 
Tevatron high-$p_{\rm T}$ physics programme.  
Dedicated measurements continue, harnessing the high statistics 
datasets: improving tau identification; testing higher-order 
calculations and PDFs and probing QCD; and making precision 
measurements of Standard Model parameters.

\begin{acknowledgments}
The author thanks the UK Science and Technology Facilities Council for financial support.
\end{acknowledgments}


\begin{thebibliography}{9}   
\bibitem{ref:cdfwzprl}
D. Acosta et al., Phys.\,Rev.\,Lett. {\bf 94} (2005) 091803; A. Aaltonen et al., J.\,Phys.\,G {\bf 34} (2007) 2457
\bibitem{ref:higgs}
A. Aaltonen et al., arXiv:0809.3930, submitted to Phys.\,Rev.\,L
\bibitem{ref:d0newZtautau}
V. Abazov et al., arXiv:0808.1306, submitted to Phys.\,Lett.\,B
\bibitem{ref:d0pt}
V. Abazov et al., Phys.\,Rev.\,Lett. {\bf 100} (2008) 102002
\bibitem{ref:cdfwwidth}
A. Aaltonen et al., Phys.\,Rev.\,Lett. {\bf 100} (2008) 071801
\bibitem{ref:pete}
P. Renton, arXiv:0804.4779
\bibitem{ref:zinviswidth}
http://www-cdf.fnal.gov/physics/ewk/2007/ZnunuWidth/

\end{thebibliography}
\end{document}